\documentclass[aps,twocolumn,pra,showpacs,floatfix]{revtex4}
\usepackage{dcolumn}

\begin{document}

\title{\bf Combination of the single-double coupled cluster 
  and the configuration interaction methods; application to barium,
  lutetium and their ions.}
\author{V. A. Dzuba}
\affiliation{School of Physics, University of New South Wales, 
Sydney 2052, Australia}
\date{\today}

\begin{abstract}
A version of the method of accurate calculations for few
valence-electron atoms which combines linearized single-double
coupled cluster method with the configuration interaction technique is
presented. The use of the method is illustrated by calculations of the
energy levels for Ba, Ba$^+$, Lu, Lu$^+$ and Lu$^{2+}$. Good agreement
with experiment is demonstrated and comparison with previous version
of the method (Safronova {\em et al}, PRA {\bf 80}, 012516 (2009)) is made.

\end{abstract} 
\maketitle
%***************************************************************************

\section{Introduction}

Many-electron atoms play an important role in studying fundamental laws
of modern physics and searching for new physics beyond the standard
model. They are used for measurements of parity and time invariance
violation in atoms (see, e. g. reviews \cite{Ginges,DF}), search for
space-time variation of fundamental constants (see, e. g. reviews
\cite{DF1,DF2}), etc. Another important area of application is the
construction of very accurate atomic \cite{aclock}) and nuclear
\cite{nclock} clocks. Planning and interpreting the measurements
require accurate atomic calculations. The calculations are also needed
to address the lack of experimental data, e.g. for such systems as
superheavy elements \cite{Pershina1,Pershina2} and highly-charged ions
\cite{Ong}. 

For atoms with one external electron above closed-shell core the best
results are achieved by the use of all-order techniques based on
either different versions of the coupled-cluster method \cite{CC1,CC2,CC3}
or correlation potential (CP) method \cite{all-order,ladder}.
For heavy atoms with several valence electrons the most accurate
methods include multi-configurational Hartree-Fock method (MCHF)
\cite{F-F} and different versions of the configuration interaction
techniques. There are also  versions of the CC method adopted for
two-valence-electron systems~\cite{Kaldor,Sahoo}.  

Configuration interaction method combined with the many-body
perturbation theory to include
core-valence correlations (the CI+MBPT method~\cite{CI+MBPT}) turned
out to be a very efficient tool for accurate calculations for
many-electron atoms having two or three valence electrons (see,
also \cite{ex1,ex2,ex3,ex4}). In this method core-valence correlations are
included into effective CI Hamiltonian in second order of the MBPT. 
Recently developed method which combines CI with the 
linearized single-double coupled cluster method~\cite{SD+CI} takes
core-valence correlations to next level of accuracy by including
certain types of the core-valence correlations to all orders. The
results for two and three valence electron atoms are significantly
better than in the CI+MBPT method~\cite{SD+CI,SD+CIres}. 

In this paper an independent version of the SD+CI method is
presented. It is very similar to what was presented in Ref.~\cite{SD+CI} 
but also has some important differences which will be discussed in
next section. We study the use of the method using barium, lutetium
and their positive ions Ba$^+$, Lu$^+$ and Lu$^{2+}$ as examples. We
demonstrate that the accuracy of the results for ions and neutral
atoms is correlated. If the SD method works well for a single-valence
electron ion then the SD+CI method should give accurate results for
corresponding neutral atom. This is important observation since it is
well known that the SD approximation does not work well for every atomic
system with one external electron (see, e.g. \cite{Johnson}) and
triple excitations often need to be included for better accuracy. 
For some atoms a good alternative to the use of single-electron SD
operator is the use of the all-order correlation
potential~\cite{all-order,ladder,Dzuba13}. 
%Therefore blind use of the SD+CI method for any two- or
%three-valence-electron atoms my lead to inaccurate results. 

%In this work we apply the SD+CI method to calculation of energy levels
%of barium, lutetium and their positive ions Ba$^+$, Lu$^+$ and
%Lu$^{2+}$. 
The results of present work for Ba are in good agreement with earlier
calculations of Ref.~\cite{SD+CI}, in spite of some differences in the
implementation of the method. We confirm that the use of the SD
approximation reduces the deviation of theoretical energies from the
experiment by about of factor of two as compared to the CI+MBPT method
in which core-valence correlations are included in second-order of
MBPT. However, better accuracy for both Ba and Ba$^+$ is achieved when
the all-order correlation potential~\cite{all-order} which includes
ladder diagrams~\cite{ladder} is used in place of the single-electron
SD operator. The two-electron operator (screening of Coulomb interaction)
is still included in the SD approximation. Deviation of theoretical 
energies from experiment are further reduced about four time for 
Ba$^+$ and about two times for Ba.

In contrast to Ba, the SD approximation produces very accurate results
simultaneously for Lu$^{2+}$, Lu$^+$ and Lu. 

\section{Method}

In the single-double linearized couple cluster method (SD) the many
electron wave function of an atom is written as an expansion over
terms containing single and double excitations of core or valence
electrons from the reference Hartree-Fock wave function into basis
states above the core (see, e.g. Ref.~\cite{CC1}). The
coefficients of the expansion are found by solving the SD equations.
The SD equations for the core have a form \cite{CC1}
\begin{eqnarray}
&&(\epsilon_a - \epsilon_m)\rho_{ma} = \sum_{bn}\tilde g_{mban}\rho_{nb} + 
 \nonumber \\
&&\sum_{bnr}g_{mbnr}\tilde\rho_{nrab}- \sum_{bcn}g_{bcan}\tilde\rho_{mnbc},
 \nonumber \\
&&(\epsilon_a+\epsilon_b-\epsilon_m-\epsilon_n)\rho_{mnab} = g_{mnab}+ 
\label{lcore} \\
&&\sum_{cd}g_{cdab}\rho_{mncd}+ \sum_{rs}g_{mnrs}\rho_{rsab} + \nonumber \\
&& \sum_r g_{mnrb}\rho_{ra}-\sum_c g_{cnab}\rho_{mc} 
+\sum_{rc}\tilde g_{cnrb}\tilde \rho_{mrac} + 
 \nonumber \\
&& \sum_r g_{nmra}\rho_{rb}-\sum_c g_{cmba}\rho_{nc} 
+\sum_{rc}\tilde g_{cmra}\tilde \rho_{nrbc}  \ \ \ \nonumber 
\end{eqnarray}

Here parameters $g$ are Coulomb integrals 
\[ g_{mnab} = \int \int \psi_m^\dagger(r_1) \psi_n^\dagger(r_2)\frac{e^2}{r_{12}}
\psi_a(r_1)\psi_b(r_2)d\mathbf{r}_1d\mathbf{r}_2, \] 
parameters $\epsilon$ are the single-electron Hartree-Fock
energies. Coefficients $\rho_{ma}$ and $\rho_{mnab}$ are the expansion
coefficients which are to be found by solving the equations
iteratively starting from
\begin{eqnarray}
&& \rho_{mnij} =
\frac{g_{mnij}}{\epsilon_i+\epsilon_j-\epsilon_m-\epsilon_n}, \label{initial}
\\
&& \rho_{ma} = 0. \nonumber
\end{eqnarray}
The tilde above $g$ or $\rho$ means the sum of direct and
exchange terms, e.g. 
\[ \tilde \rho_{nrbc} = \rho_{nrbc} - \rho_{nrcb}. \]
Indexes $a,b,c$ numerate states in atomic core, indexes
$m,n,r,s$ numerate states above the core, indexes $i,j$ numerate
any states.

The correction to the energy of the core
\begin{equation}
\delta E_C = \frac{1}{2}\sum_{mnab} g_{abmn}\tilde \rho_{nmba}
\end{equation}
is used to control the convergence.

In the case of single electron above closed-shell core
the SD equations for a particular valence state $v$ can be obtained from
(\ref{lcore}) by replacing index $a$ by $v$ and replacing $\epsilon_a$ by
$\epsilon_v + \delta\epsilon_v$ where
\begin{equation}
  \delta \epsilon_v = \sum_{mab}g_{abvm}\tilde\rho_{mvab}+
  \sum_{mnb}g_{vbmn}\tilde\rho_{mnvb} 
\label{eq:sdv}
\end{equation}
is a correction to the energy of the valence electron.
The SD equations are solved iteratively first for the core and than
for as many valence states $v$ as needed. 

In the case of more than one valence electron above closed-shell core
interaction between valence electrons needs to be included. This can
be done with the use of the configuration interaction (CI) technique.
To combine the CI technique with the SD method one needs to modify the
SD equations for valence states. The SD equations for the
core (\ref{lcore}) remain the same. To see why and how the equations
should be modified it is instructive to consider an example -- the use
of the CI technique for an atom with one external electron above
closed shells. The result of the CI calculations for such system must
be equivalent to the SD calculations and the resulting energy of the
calculated state should be very close to those given by
Eq.~(\ref{eq:sdv}). However, no valence state is treated in the CI
approach as an initial approximation. Instead, all single electron
states of the same symmetry are treated as a basis and the wave
function of the valence electron is presented as an expansion
\begin{equation}
\psi_v =\sum_n c_n \psi_n.
\label{eq:ci1}
\end{equation}
Here $\psi_n$ are single-electron basis states lying above the core,
expansion coefficients $c_n$ and the energy of the valence state $v$
are to be found via matrix diagonalization. The SD equations should be
run for every basis state $\psi_n$ in the expansion. However, the
energy parameter in all these SD equations must be the same and close
to the energy of the state which is to be found. If the lowest state
of the given symmetry is to be found then the natural choice is to use
the Hartree-Fock energy of the lowest valence basis state.

Another modification comes from the need to exclude double
counting. All terms in the SD equations which have only excitations of
valence electrons must be removed since valence excitations are
included in the CI calculations.  

The modified SD equations for valence states have the form
\begin{eqnarray}
&&(\epsilon_0 - \epsilon_m)\rho_{mv} = \sum_{bn}\tilde g_{mban}\rho_{nb} + 
 \nonumber \\
&&\sum_{bnr}g_{mbnr}\tilde\rho_{nrvb}- \sum_{bcn}g_{bcvn}\tilde\rho_{mnbc},
 \nonumber \\
&&(\epsilon_0+\epsilon_b-\epsilon_m-\epsilon_n)\rho_{mnvb} = g_{mnvb}+ 
\label{lval} \\
&&\sum_{cd}g_{cdvb}\rho_{mncd}+ \sum_{rs}g_{mnrs}\rho_{rsvb} - \nonumber \\
&& \sum_c g_{cnvb}\rho_{mc} 
+\sum_{rc}\tilde g_{cnrb}\tilde \rho_{mrvc} +  \nonumber \\
&& \sum_r g_{nmrv}\rho_{rb} -\sum_c g_{cmbv}\rho_{nc} 
+\sum_{rc}\tilde g_{cmrv}\tilde \rho_{nrbc}.   \nonumber 
\end{eqnarray}
These equations are obtained from (\ref{lcore}) by replacing core
index $a$ by valence index $v$, removing the term $\sum_r
g_{mnrb}\rho_{rv}$ which has only valence excitations, and replacing
$\epsilon_a$ by $\epsilon_0$. The energy parameter $\epsilon_0$ is
fixed and is the same for all states in the expansion
(\ref{eq:ci1}). Usually it is chosen to be the Hartree-Fock energy of
the lowest basis state of given symmetry. Note that expression
(\ref{eq:sdv}) is not used in the CI calculations. Instead, the energy
of the valence state is found as an eigenstate of the CI matrix. This
approach is very similar to one used in Ref.~\cite{positron} for
positron binding to atoms. It can also be used for negative
ions. Neither positron nor extra electron are bound to an atom in
the Hartree-Fock approximation. Therefore, their states above the core
cannot be used as initial approximation but should be used as basis
states for the CI calculations. 

In case of more than one external electron, the Coulomb interaction
between valence electrons needs to be modified as well. Replacing in
the equations for the double excitation coefficients (\ref{lcore})
core indexes $a,b$ by valence indexes $v,w$ and removing terms which
have only valence excitations we get the expressions for screened
Coulomb integrals to be used in the CI calculations  
\begin{eqnarray}
&&q_{mnvw} = g_{mnvw}+ \nonumber \\
&&\sum_{cd}g_{cdvw}\rho_{mncd} -  \sum_c \left( g_{cnvw}\rho_{mc} 
+ g_{cmwv}\rho_{nc} \right) + \nonumber \\
&&\sum_{rc} \left( g_{cnrw}\tilde \rho_{mrvc} 
+ g_{cmrv}\tilde \rho_{nrwc}  + g_{cnwr} \rho_{mrvc}
\right. \label{qscreen} \\   
&&+ g_{cmvr} \rho_{nrwc} - g_{cmwr} \rho_{nrcv} - \left. g_{cnvr}
  \rho_{mrcw} \right)  \nonumber  
\end{eqnarray}
The effective CI Hamiltonian can be written as a sum of one and two
electron parts
\begin{equation}
\hat H^{\rm CI} = \sum_i^{N_v} \hat h_1(r_i) + \sum_{i<j}^{N_v} \hat
h_2(r_i,r_j). 
\label{eq:CI}
\end{equation}
Here $N_v$ is the number of valence electrons. The single electron part is
given by
\begin{equation}
\hat h_1 = c {\mathbf \alpha} {\mathbf p} + (\beta -1 )mc^2 +
V_{core} + \hat \Sigma_1,
\label{eq:h1}
\end{equation}
where ${\mathbf \alpha}$ and $\beta$ are Dirac matrixes, $V_{core}$ is
the self-consistent potential of the atomic core (including nuclear
part), $\hat \Sigma_1$ is the single-electron correlation operator
responsible for the correlation interaction of a valence electron with
the core. Its matrix elements are obtained from (\ref{lval}) and can 
be written as 
\begin{equation}
\langle v |\hat \Sigma_1| m \rangle = (\epsilon_0 -
\epsilon_m)\rho_{mv}.
\label{eq:sigma1}
\end{equation}
The two-electron part $\hat h_2$ of the CI Hamiltonian (\ref{eq:CI})
is the sum of the Coulomb interaction and the two-electron correlation
operator $\hat \Sigma_2$. Matrix elements of $\hat h_2$ in the SD
approximation are given by (\ref{qscreen}). 

The notation $\hat \Sigma$ for the operator of the core-valence
correlations was introduced in Ref.~\cite{CI+MBPT} in the framework of
the CI method combined with the many-body perturbation theory
(MBPT). In this and following works~\cite{ex1,ex2,ex3,ex4} the $\hat
\Sigma_1$ and $\hat \Sigma_2$ operators were calculated in the lowest
second-order of the MBPT. This corresponds to substituting the initial
approximation (\ref{initial}) into (\ref{lval}) and (\ref{qscreen})
without iterating the SD equations (\ref{lcore},\ref{lval}). Thus, present work
further advances the method by including higher-order
correlations. This leads to significant improvement in the accuracy of
the calculations for many atomic systems.

\label{s:method}

The SD+CI method was first developed in Ref.~\cite{SD+CI}. Its 
implementation in present work is independent and slightly
different. Most of the difference is in energy denominators. Energy
denominators in equations (\ref{lval}) and (\ref{qscreen}) are hidden
in the expressions for the excitation coefficients $\rho$. In
corresponding expressions of Ref.~\cite{SD+CI} (see Eq.(22-24) in
\cite{SD+CI}) the energy denominators are shown explicitly and they
are different from what can be fund in  (\ref{lval}) and
(\ref{qscreen}). Energy denominator of every term containing valence
state $v$ is corrected in \cite{SD+CI} by the energy difference 
$\tilde \epsilon_v - \epsilon_v$, where $\epsilon_v$ is the 
Hartree-Fock energy of the
valence state $v$ and $\tilde \epsilon_v$ is an external parameter. It
can be chosen to be the energy of the lowest state of given symmetry
or it can be used as a fitting parameter. Let us consider in more
detail why different energy denominators may appear.

Moving core excitations into valence space leads to the following
second-order corrections to the matrix elements of the CI Hamiltonian
\cite{CI+MBPT} 
\begin{equation}
\Delta H^{\rm CI}_{IJ} = \sum_M \frac{\langle I|U|M \rangle \langle
  M|U|J \rangle}{E - E_M},
\label{B-W}
\end{equation}
where $U$ is residual Coulomb interaction, $|I\rangle$ and $|J\rangle$ are
many-electron states in the valence space, states $|M\rangle$ are many
electron states with excitations from the core, $E_M$ is the
many-electron energy of the state $|M\rangle$, $E$ is the energy of
the state to be found in the calculations. Expression (\ref{B-W})
corresponds to the Brillouin-Wigner (BW) version of the MBPT.
The alternative is to use the Rayleigh-Schr\"{o}dinger (RS) MBPT in which
Eq. (\ref{B-W}) transfers to
\begin{equation}
\Delta H^{\rm CI}_{IJ} = \sum_M \frac{\langle I|U|M \rangle \langle
  M|U|J \rangle}{E_I - E_M}.
\label{B-W}
\end{equation}
There are two reasons to use the BW MBPT rather than the RS one. First, the CI
matrix is not symmetric in the RS approach, $H^{\rm CI}_{IJ} \neq
H^{\rm CI}_{JI}$. Second, energy denominator $E_I - E_M$ for highly
excited state $|I\rangle$ may accidently become very small leading to
unphysical enhancement of otherwise small contribution. 

The downside of the use of the BW version of the MBPT is that there is
no exact solution to the problem of reducing the many-electron
expression (\ref{B-W}) to single- and two-electron operators $\hat
\Sigma_1$ and $\hat \Sigma_2$. Thus different approximations can be
used. 

One can see from (\ref{B-W}) that the core-valence correlation
operator $\hat \Sigma$ is the energy-dependent operator which should
be calculated at the energy close to the energy of the state of
interest. The  $\tilde \epsilon_v - \epsilon_v$ corrections to the
energy denominators used in Ref.~\cite{SD+CI} were introduced to
ensure correct dependence of the $\hat \Sigma$ operator on the energy
(see also \cite{Kozlov}). This might be important for highly excited
states. For low lying states which are always of the most interest the
corrections are less important and can be neglected. 

\subsection{Basis states}

A complete set of single-electron states is needed for solving the SD
equations (\ref{lcore},\ref{lval}) and for construction of the
many-electron states for the CI calculations. We use the same B-spline
technique~\cite{JohSap86} for both purposes. Forty B-spline states
of the order of nine are calculated in a box of radius 40 $a_B$ in
each partial wave up to $l_{\rm max}=6$. All of them are used for
calculating terms in the SD equations indexes $n,m,r,s$ in
(\ref{lcore},\ref{lval},\ref{qscreen},\ref{eq:sigma1})). The SD
equation for valence states are solved for few (three of four) states
above the core in each partial wave up to $l_{\rm max}=3$. The
second-order correlation potential $\Sigma$ is used for higher
states. Fourteen states above the core in each partial wave up to
$l_{\rm max}=4$ are used in the CI calculations. With this choice of
the parameters the basis is sufficiently saturated.

\subsection{Breit and QED corrections}

Breit and quantum electrodynamic (QED) corrections are not very important
for barium and lutetium. We include them however to be sure that
remaining deviation of the calculated energies from experiment is
mostly due to higher-order correlations. We do this in the same way as
in our previous works (see, e.g. \cite{Breit,Dzuba13}).

We treat Breit interaction in zero energy transfer approximation. The
Breit Hamiltonian includes magnetic interaction and retardation:
\begin{equation}
\hat H^{B}=-\frac{\mbox{\boldmath$\alpha$}_{1}\cdot \mbox{\boldmath$\alpha$}_{2}+
(\mbox{\boldmath$\alpha$}_{1}\cdot {\bf n})
(\mbox{\boldmath$\alpha$}_{2}\cdot {\bf n})}{2r} \ .
\label{Breit}
\end{equation}  
Here ${\bf r}={\bf n}r$, $r$ is the distance between electrons, and 
$\mbox{\boldmath$\alpha$}$ is the Dirac matrix.

Breit interaction is included in the self-consistent Hartree-Fock
procedure. Thus the effect of Breit interaction on self consistent
atomic potential is included. This effect, which is often called the
relaxation effect, is still linear in Breit but non-linear in Coulomb
interaction. Its inclusion leads to more accurate results than
treating Breit interaction perturbatevely.

%\subsection{QED corrections}

To include the QED corrections we use the radiative potential approach
developed in Ref.~\cite{radpot}. This radiative potential has the form
\begin{equation}
V_{\rm rad}(r)=V_U(r)+V_g(r)+ V_e(r) \ ,
\end{equation}
where $V_U$ is the Uehling potential, $V_g$ is the potential arising from the 
magnetic formfactor, and $V_e$ is the potential arising from the
electric formfactor. The $V_U$ and $V_e$ terms can be considered as
additions to nuclear potential while inclusion of $V_g$ leads to some
modification of the Dirac equation (see Ref.~\cite{radpot} for details).

\section{Results}

Performing calculations in the $V^{N-2}$ approximation for Ba and
$V^{N-3}$ approximation for Lu we have pairs of atomic systems with
similar electron structure. Ba has electron structure similar to
Lu$^+$ and Ba$^+$ is similar to Lu$^{2+}$. It is natural to expect
some similarities in the results. It turns out however that barium and
lutetium are sufficiently different so that best results are achieved
with slightly different procedures. We will therefore discuss them
separately. 

\subsection{Barium and its positive ion}

\begin{table}
\caption{Removal energies (cm$^{-1}$) of the lowest $s,p,d$ states of
  Ba$^+$ in SD+CI and CP+CI approximations.
  $\Delta = E_{\rm theor} - E_{\rm expt}$.}
\label{t:ba+}
\begin{ruledtabular}
\begin{tabular}{l rrrrr}
\multicolumn{1}{c}{State} &
\multicolumn{1}{c}{SD+CI} &
\multicolumn{1}{c}{$\Delta$} &
\multicolumn{1}{c}{CP+CI} &
\multicolumn{1}{c}{$\Delta$} &
\multicolumn{1}{c}{Expt.\footnotemark[2]} \\
\hline
$6s_{1/2}$ &   81210  & 524  &  80719  &   33  &   80686 \\
$5d_{3/2}$ &   76341  & 529  &  75969  &  157  &   75812 \\
$5d_{5/2}$ &   75474  & 463  &  75108  &   97  &   75011 \\
$6p_{1/2}$ &   60710  & 286  &  60512  &   88  &   60424 \\
$6p_{3/2}$ &   58980  & 246  &  58794  &   60  &   58734 \\
\end{tabular}
\footnotetext[1]{Ref.~\cite{NIST}.}
\end{ruledtabular}
\end{table}

Table \ref{t:ba+} presents results of calculations of the energy
levels of Ba$^+$. As one can see the accuracy of the SD approximation
for the ion is on the level 0.5 - 0.7\%. Significantly better accuracy
can be achieved if the all-order correlation
potential~\cite{all-order,ladder}  is used~\cite{Dzuba13}. 
Corresponding numbers are presented in Table
\ref{t:ba+} under the header ``CP+CI''. Note that these numbers are
slightly different from those presented in Ref.~\cite{Dzuba13}. The
difference is due to the fact that in present work the energies were
obtained as a result of matrix diagonalization while in
Ref.~\cite{Dzuba13} they were found by solving single-electron equations
for Brueckner orbitals. 

Similar improvement in accuracy of the
calculations can be achieved for neutral barium if single-electron SD
matrix elements (\ref{eq:sigma1}) are replaced by the matrix elements
of the all-order correlation potential as in Ref.~\cite{Dzuba13}.

The results for neutral barium are presented in Table \ref{t:ba}. The
results of previous calculations in the SD+CI
approximation~\cite{SD+CI} are also presented for comparison. As one
can see, in spite of some difference in the methods (see discussion of
energy denominators in section \ref{s:method}), the results of both
SD+CI calculations are very close to each other. It was pointed out in
Ref.~\cite{SD+CI} that the deviation of the calculated energies from
experiment in the SD+CI calculations is about two times smaller than
in the CI+MBPT calculations. This is the effect of selected
higher-order core valence correlations included in the SD+CI method
but not in the CI+MBPT method. Which higher-order correlations play
the most important role varies from atom to atom. The SD approximation
is not always the best choice. It turns out that for barium better
results are achieved when single-electron SD
matrix elements (\ref{eq:sigma1}) are replaced by the matrix elements
of the all-order correlation potential, while screened Coulomb
integrals are kept the same (see Eq. (\ref{qscreen})). Corresponding
results are presented in Table \ref{t:ba} in the CP+CI column. As one
can see, the difference with the experimental energies is further
reduced by about two times.

The accuracy of present calculations for barium is better than in
previous calculations with the use of the CI+MBPT 
method~\cite{ex1,Ba1,Ba2}. This is in spite of the fact that present
calculations are pure {\it ab initio} ones while in earlier calculations 
rescaling of the second-order correlation operator $\hat \Sigma$ was
used to fit the energies of the lowest states.

\begin{table}
\caption{Excitation energies (cm$^{-1}$) of the lowest states of
  Ba in SD+CI and CP+CI approximations;
  $\Delta = E_{\rm theor} - E_{\rm expt}$.}
\label{t:ba}
\begin{ruledtabular}
\begin{tabular}{llc rrr rrrr}
\multicolumn{2}{c}{State} &
\multicolumn{1}{c}{$J$} &
\multicolumn{2}{c}{Ref.~\cite{SD+CI}} &
\multicolumn{4}{c}{This work} & \\

&&&\multicolumn{1}{c}{SD+CI} &
\multicolumn{1}{c}{$\Delta$} &
\multicolumn{1}{c}{SD+CI} &
\multicolumn{1}{c}{$\Delta$} &
\multicolumn{1}{c}{CP+CI} &
\multicolumn{1}{c}{$\Delta$} &
\multicolumn{1}{c}{Expt.\footnotemark[1]} \\
\hline
$6s^2$ & $^1$S     & 0 &     0 &     &     0 &      &     0 &     &     0 \\
$6s5d$ & $^3$D     & 1 &  9249 & 216 &  8882 & -151 &  8936 & -97 &  9033 \\
       &           & 2 &  9441 & 225 &  9132 &  -84 &  9187 &  71 &  9216 \\
       &           & 3 &  9840 & 243 &  9505 &  -92 &  9560 & -37 &  9597 \\
$6s5d$ & $^1$D     & 2 & 11721 & 326 & 11471 &   76 & 11508 & 113 & 11395 \\
$6s6p$ & $^3$P$^o$ & 0 & 12556 & 290 & 12541 &  275 & 12325 &  59 & 12266 \\
       &          & 1 & 12919 & 282 & 12898 &  261 & 12679 &  42 & 12637 \\
       &          & 2 & 13819 & 304 & 13796 &  281 & 13568 &  53 & 13515 \\
$6s6p$ & $^1$P$^o$ & 1 & 18292 & 232 & 18173 &  113 & 17973 & -87 & 18060 \\
$5d^2$ & $^3$F     & 2 &       &     & 20722 & -212 & 20850 & -84 & 20934 \\
       &           & 3 &       &     & 20956 & -294 & 21080 &-170 & 21250 \\
       &           & 4 &       &     & 21462 & -162 & 21584 & -40 & 21624 \\
$5d6p$ & $^3$F$^o$ & 2 &       &     & 22154 &   89 & 22006 & -59 & 22065 \\
       &          & 3 &       &     & 23050 &  103 & 22916 & -31 & 22947 \\
       &          & 4 &       &     & 23912 &  155 & 23768 &  11 & 23757 \\
$5d^2$ & $^1$D    & 2 &       &     & 22931 & -131 & 23062 &   0 & 23062 \\
$5d6p$ & $^1$D$^o$& 2 &       &     & 23237 &  163 & 23035 & -39 & 23074 \\
$5d^2$ & $^3$P   & 0 &       &     & 22729 & -480 & 22862 &-212 & 23209 \\
       &         & 1 &       &     & 22877 & -603 & 23018 &-462 & 23480 \\
       &         & 2 &       &     & 23663 &  255 & 23794 &-124 & 23918 \\
$5d6p$ & $^3$D$^o$&1 &       &     & 24266 &   74 & 24044 &-148 & 24192 \\
       &         & 2 &       &     & 24635 &  103 & 24410 &-122 & 24532 \\
       &         & 3 &       &     & 25110 &  130 & 24885 & -95 & 24980 \\
$5d6p$ & $^3$P$^o$& 0 &       &     & 25721 &   79 & 25528 &-114 & 25642 \\
       &         & 1 &       &     & 25789 &   85 & 25600 &-104 & 25704 \\
       &         & 2 &       &     & 26088 &  131 & 25902 & -55 & 25957 \\
$6s7s$ & $^3$S   & 1 &       &     & 26425 &  265 & 26135 & -25 & 26160 \\
%$5d^2$ & $^1$S   & 0 &       &     & 25648 & -512 & 25754 &-1003& 26757 \\
%$5d6p$ & $^1$F$^o$&3 &       &     & 26959 &  143 & 26917 & 101 & 26816 \\
%$6s7s$ & $^1$S   & 0 &       &     & 28463 &  233 & 28265 &  35 & 28230 \\
%$5d6p$ & $^1$P$^o$&1 &       &     & 28787 &  233 & 28547 &  -7 & 28554 \\
\end{tabular}
\footnotetext[1]{Ref.~\cite{NIST}.}
\end{ruledtabular}
\end{table}

\subsection{Lutetium and its ions}

\begin{table}
\caption{Removal energies (cm$^{-1}$) of the lowest $s,p,d$ states of
  Lu$^{2+}$ in SD and SD+CI approximations.
  $\Delta = E_{\rm theor} - E_{\rm expt})$.}
\label{t:Lu++}
\begin{ruledtabular}
\begin{tabular}{l rrrrr}
\multicolumn{1}{c}{State} &
\multicolumn{1}{c}{SD} &
\multicolumn{1}{c}{$\Delta$} &
\multicolumn{1}{c}{SD+CI} &
\multicolumn{1}{c}{$\Delta$} &
\multicolumn{1}{c}{Expt.\footnotemark[1]} \\
\hline
$6s_{1/2}$ &  169705  & 691  & 169723  &  709  & 169014 \\
$5d_{3/2}$ &  163324  &  18  & 163241  &  -65  & 163306 \\
$5d_{5/2}$ &  160304  & -62  & 160204  & -162  & 160366 \\
$6p_{1/2}$ &  130956  & 343  & 131017  &  404  & 130613 \\
$6p_{3/2}$ &  124558  & 249  & 124587  &  278  & 124309 \\
\end{tabular}
\footnotetext[1]{Ref.~\cite{NIST}.}
\end{ruledtabular}
\end{table}

The results for energy levels of Lu$^{2+}$ are presented in Table
\ref{t:Lu++}. There are two sets of the SD results. One (called
SD) corresponds to the standard SD method for atoms with one external
electron in which the SD equations are iterated for a specific valence
state and correction to the energy is calculated using
(\ref{eq:sdv}). Another set (called SD+CI) is based on the CI
method. Hartree-Fock valence states are treated as a basis (as in
Eq. (\ref{eq:ci1})) and the energy of the valence state is found as an
eigenstate of the CI Hamiltonian. This two approaches are almost
equivalent. Some small difference in results is mostly due to the fact
that equations (\ref{lval}) used in the CI calculations are iterated
with the fixed energy parameter $\epsilon_0$ while energy of the
valence state (\ref{eq:sdv}) in the standard SD method changes on
every iteration. Other factor which can make minor contribution to the
difference is incompleteness of the basis in the expansion
(\ref{eq:ci1}) and the fact that the SD equations (\ref{lval}) are
iterated only for few first basis states in the expansion.
In the end the difference between SD and SD+CI results is very
small. The difference of theses result and experiment is also
small. It is on the level 0.4\% or better. This reflects general
trend for isoelectronic sequences of alkali atoms. For example, the
SD approximation gives poor accuracy for cesium~\cite{Johnson}, better
accuracy for Ba$^+$ (see Table \ref{t:ba+}) and even better accuracy
for Lu$^{2+}$. In contrast to Ba$^+$, using the correlation potential
method does not lead to improvement in accuracy. Therefore we limit
further calculations to the SD+CI method.

Tables \ref{t:Lu+} and \ref{t:Lu} show the results of the SD+CI
calculations for the energy levels of Lu$^+$ and Lu. The results of
previous similar calculations for Lu are also shown for comparison.

\begin{table}
\caption{Excitation energies (cm$^{-1}$) of the lowest states of
  Lu$^+$ in SD+CI; 
  $\Delta = E_{\rm theor} - E_{\rm expt}$.}
\label{t:Lu+}
\begin{ruledtabular}
\begin{tabular}{llc rrr}
\multicolumn{2}{c}{State} &
\multicolumn{1}{c}{$J$} &
\multicolumn{1}{c}{SD+CI} &
\multicolumn{1}{c}{$\Delta$} &
\multicolumn{1}{c}{Expt.\footnotemark[1]} \\
\hline
$6s^2$ & $^1$S     & 0 &     0 &  0   &     0     \\
$6s5d$ & $^3$D     & 1 & 11948 & 152 & 11796  \\
       &           & 2 & 12695 & 260 & 12435  \\
       &           & 3 & 14473 & 274 & 14199  \\
$6s5d$ & $^1$D     & 2 & 17892 & 560 & 17332  \\
                                              
$6s6p$ & $^3$P$^o$ & 0 & 27657 & 393 & 27264  \\
       &           & 1 & 28891 & 388 & 28503  \\
       &           & 2 & 32918 & 465 & 32453  \\

$6s6p$ & $^1$P$^o$ & 1 & 18292 & 232 & 18060 \\

$5d^2$ & $^3$F     & 2 & 29751 & 345 & 29406 \\
       &           & 3 & 31238 & 349 & 30889 \\
       &           & 4 & 32985 & 482 & 32503 \\
                                            
$5d^2$ & $^3$P     & 0 & 35673 &  21 & 35652 \\
       &           & 1 & 36574 &  17 & 36557 \\
       &           & 2 & 39202 & 628 & 38574 \\
                                            
$5d^2$ & $^1$D     & 2 & 36563  & 465 & 36098 \\

$5d6p$ & $^3$F$^o$ & 2 & 41789 &  565 & 41224 \\
       &           & 3 & 45575 &  657 & 44918 \\
       &           & 4 & 49304 &  739 & 48536 \\

$5d6p$ & $^1$D$^o$ & 2 & 46149 & 691 & 45458 \\

$5d6p$ & $^3$D$^o$ & 1 &  46015 & 483 & 45532 \\
       &           & 2 &  47474 & 570 & 46904 \\
       &           & 3 &  49359 & 626 & 48733 \\
                                             
$5d6p$ & $^3$P$^o$ & 0 &  50505 & 542 & 49963 \\
       &           & 1 &  50616 & 567 & 50049 \\
       &           & 2 &  51884 & 683 & 51201 \\
\end{tabular}
\footnotetext[1]{Ref.~\cite{NIST}.}
\end{ruledtabular}
\end{table}

\begin{table}
\caption{Excitation energies (cm$^{-1}$) of the lowest states of
  Lu in SD+CI approximation; comparison with previous work and experiment. 
  $\Delta = E_{\rm theor} - E_{\rm expt}$.}
\label{t:Lu}
\begin{ruledtabular}
\begin{tabular}{llc rrrrr}
\multicolumn{2}{c}{State} &
\multicolumn{1}{c}{$J$} &
\multicolumn{2}{c}{Ref.~\cite{Safronova}} &
\multicolumn{2}{c}{This work} & \\
&&&\multicolumn{1}{c}{Energy} &
\multicolumn{1}{c}{$\Delta$} &
\multicolumn{1}{c}{Energy} &
\multicolumn{1}{c}{$\Delta$} &
\multicolumn{1}{c}{Expt.\footnotemark[1]} \\
\hline
$5d6s^2$  & $^2$D     &  3/2 &     0 &      &     0 &      &     0 \\  
          &           &  5/2 &  2014 &   20 &  2123 &  129 &  1994 \\  
                                                          
$6s^26p$  & $^2$P$^o$ &  1/2 &  3910 & -226 &  3975 & -161 &  4136 \\  
          &           &  3/2 &  7228 & -249 &  7367 & -109 &  7476 \\  
                                                          
$5d6s6p$  & $^4$F$^o$ &  3/2 & 17723 &  296 & 17502 &   75 & 17427 \\  
          &           &  5/2 & 18789 &  285 & 18652 &  147 & 18505 \\  
          &           &  7/2 & 20731 &  299 & 20596 &  163 & 20433 \\  
          &           &  9/2 & 22911 &  302 & 22786 &  177 & 22609 \\  
                                                          
$5d^26s$  & $^4$F     &  3/2 & 19182 &  331 & 18662 & -189 & 18851 \\  
          &           &  5/2 & 19737 &  334 & 19248 & -155 & 19403 \\  
          &           &  7/2 & 20578 &  331 & 20095 & -152 & 20247 \\  
          &           &  9/2 & 21591 &  349 & 21159 &  -83 & 21242 \\  
                                                          
$5d6s6p$  & $^4$D$^o$ &  1/2 & 20995 &  233 & 20795 &   33 & 20762 \\  
          &           &  3/2 & 21448 &  253 & 21284 &   89 & 21195 \\  
          &           &  5/2 & 22504 &  283 & 22352 &  130 & 22222 \\  
          &           &  7/2 & 23795 &  271 & 23665 &  141 & 23524 \\  
                                                          
$5d6s6p$  & $^2$D$^o$ &  3/2 & 22376 &  252 & 22534 &  410 & 22125 \\  
          &           &  5/2 & 21735 &  273 & 21775 &  313 & 21462 \\  
                                                          
$5d^26s$  & $^4$P     &  5/2 & 23242 &  440 & 23084 &  282 & 22802 \\  
          &           &  1/2 & 21860 &  388 & 21803 &  331 & 21472 \\  
          &           &  3/2 & 22849 &  382 & 22671 &  203 & 22468 \\  
                                                          
$5d6s6p$  & $^4$P$^o$ &  1/2 & 24520 &  412 & 24262 &  153 & 24109 \\  
          &           &  3/2 & 24786 &  478 & 24563 &  255 & 24308 \\  
          &           &  5/2 & 25774 &  583 & 25524 &  332 & 25192 \\  
                                                          
$6s^27s$  & $^2$S     &  1/2 &       &      & 25408 &  890 & 24126 \\  
                                                          
$5d^26s$  & $^2$D     &  3/2 & 25015 &  497 & 25128 & -390 & 24518 \\  
          &           &  5/2 &       &      & 25162 & -549 & 24711 \\  
\end{tabular}
\footnotetext[1]{Ref.~\cite{NIST}.}
\end{ruledtabular}
\end{table}

\section{Conclusion}

A version of the method of calculations for many-electron atoms
which combines configuration interaction with the single-double
linearized coupled cluster approach is presented. This version is 
simpler than previous one but gives results on the same level
of accuracy. The accuracy is better than in the widely used
CI+MBPT method. It can be further improved with a different choice
of the single-electron correlation operator $\hat \Sigma$.  
The best choice for barium is the all-order correlation potential
which was widely used before for the systems with one external
electron above closed shells. It is demonstrated that the accuracy
of calculation for neutral atoms and positive ions of these atoms
is correlated. Therefore, one can  choose  an adequate 
approximation for the ions first before proceeding to neutral
atoms. This is very convenient since calculations for ions take
much less computer resources. In the end, we have another method
with can be widely used for accurate calculations in many atomic
systems.

\begin{acknowledgments}

The author is grateful to M. G. Kozlov for useful discussions.
The work is partly supported by the Australian Research Council.

\end{acknowledgments}

\end{document}